\documentstyle[12pt]{article} \textheight 23cm \textwidth 15cm
\def\1{{\chi}}

\begin{document}
\title {{Unified $(r,s)$-relative entropy}\thanks{This project is supported by Natural Science Foundation of China (10771191 and
10471124) and Natural Science Foundation of Zhejiang Province of
China (Y6090105).}}
\author {Wang Jiamei, Wu Junde\date{}\thanks{Department of
Mathematics, Zhejiang University, Hangzhou 310027, P. R. China.
E-mail: wjd@zju.edu.cn}} \maketitle
\begin{abstract} {\noindent In this paper, we introduce and study unified $(r,s)$-relative entropy and quantum unified $(r,s)$-relative
entropy, in particular, our main results of quantum unified
$(r,s)$-relative entropy are established on the infinite dimensional
separable complex Hilbert spaces.}
\end{abstract}

\vskip 0.2in\noindent


\vskip0.2in

\noindent {\bf 1. Introduction}

\vskip0.2in

\noindent In 1991, Rathie and Taneja introduced the unified
$(r,s)$-entropy which generalized many classical entropies ([1]),
that is, let $A=(a_{1},a_{2},\cdots,a_{n})$ be a discrete
probability distribution satisfies that $0<a_{i}\leq 1$ and
$\sum_{i=1}^{n}a_{i}=1$. If we denote
$p(r)=\sum_{i=1}^{n}a_{i}^{r}$, then for any $r> 0$ and any real
number $s$, the unified $(r,s)-$entropy is defined by
\begin{eqnarray*}E_{r}^{s}(A)= \left \{
\begin{array}{ll}
H_{r}^{s}(A), & {\rm if\ } r\neq 1,s\neq 0,\\
H_{r}(A), & {\rm if\ } r\neq 1,s = 0,\\
H^{r}(A), & {\rm if\ } r\neq 1,s = 1,\\
_{\frac{1}{r}}H(A), & {\rm if\ } r\neq 1,s = 1/r,\\
H(A), & {\rm if\ } r= 1,\end{array} \right.
\end{eqnarray*}
where $$H_{r}^{s}(A)=[(1-r)s]^{-1}[p(r)^{s}-1],$$ $$
H_{r}(A)=(1-r)^{-1}\ln p(r),$$$$H^{r}(A)=(1-r)^{-1}(p(r)-1),$$$$
_{r}H(A)=(r-1)^{-1}[p(\frac{1}{r})^{r}-1],$$$$
H(A)=-\sum_{i=1}^{n}a_{i}\ln {a_{i}}$$
 are the $(r, s)$-entropy, R$\acute{e}$nyi entropy of order $r$, the Tsallis entropy, the entropy of type $r$ and the well-known Shannon entropy,
respectively.

\vskip0.1in

In 2006, Hu and Ye introduced the quantum version of the unified
$(r,s)$-entropy ([2]), that is, let $H$ be a complex Hilbert space
and $\rho$ a state (see [3]) on $H$. If we denote $P(r)=tr(\rho^r)$,
then for any $r>0$ and any real number $s$, the quantum unified
$(r,s)-$entropy is defined by
\begin{eqnarray*}E_{r}^{s}(\rho)= \left \{
\begin{array}{ll}
S_{r}^{s}(\rho), & {\rm if\ } r\neq 1,s\neq 0,\\
S_{r}(\rho), & {\rm if\ } r\neq 1,s = 0,\\
S^{r}(\rho), & {\rm if\ } r\neq 1,s = 1,\\
_{\frac{1}{r}}S(\rho), & {\rm if\ } r\neq 1,s = 1/r,\\
S(\rho), & {\rm if\ } r= 1,\end{array} \right.
\end{eqnarray*}
where $$S_{r}^{s}(\rho)=[(1-r)s]^{-1} \left[P(r)^{s}-1 \right],$$$$
 S_{r}(\rho)=(1-r)^{-1}\ln P(r),$$$$
S^{r}(\rho)=(1-r)^{-1} \left(P(r)-1 \right),$$$$
_{r}S(\rho)=(r-1)^{-1}\left[P(\frac{1}{r})^{r}-1
\right],$$$$S(\rho)=-tr(\rho\ln \rho)$$ are the quantum $(r,
s)$-entropy, the quantum R$\acute{e}$nyi entropy of order $r$, the
quantum Tsallis entropy, the quantum entropy of type $r$ and the
well-known Von Neumann entropy, respectively.

\vskip0.1in

On the other hand, although the  R$\acute{e}$nyi relative entropy of
order $r$ ([4]), the Tsallis relative entropy of degree $r$ (([5]),
the relative entropy ([3]), even the quantum R$\acute{e}$nyi
relative entropy ([4]) and quantum Tsallis relative entropy of
degree $r$ ([5-6]) were studied, respectively, nevertheless, until
now, we do not find the works of unified $(r,s)$-relative entropy
and quantum unified $(r,s)$-relative entropy. In this paper, we fill
this gap.

\vskip0.2in

\noindent {\bf 2. The unified $(r,s)$-relative entropy}

\vskip0.2in

\noindent Let $A=(a_{1},a_{2},\cdots,a_{n})$,
$B=(b_{1},b_{2},\cdots,b_{n})$ be two discrete probability
distributions satisfying $0< a_{i},b_{i}< 1$ and
$\sum\limits_{i=1}^{n}a_{i}=\sum\limits_{i=1}^{n}b_{i}=1$. Then for
any $r> 0$ and any real number $s$, the unified $(r,s)-$relative
entropy is defined by
\begin{eqnarray*}E_{r}^{s}(A\|B)= \left \{
\begin{array}{ll}
H_{r}^{s}(A\|B), & {\rm if\ } r\neq 1,s\neq 0,\\
H_{r}(A\|B), & {\rm if\ } r\neq 1,s = 0,\\
H^{r}(A\|B), & {\rm if\ } r\neq 1,s = 1,\\
_{\frac{1}{r}}H(A||B), & {\rm if\ } r \neq 1, s =\frac{1}{r},\\
 H(A\|B), & {\rm if\ } r=1,
 \end{array} \right.
\end{eqnarray*}
where
\begin{eqnarray*}H_{r}^{s}(A\|B)=-[(1-r)s]^{-1}\left[\left(\sum_{i=1}^{n}a_{i}\frac{a_{i}^{r-1}}{b_{i}^{r-1}}
\right)^{s}-1\right],r> 0,r\neq 1,s\neq 0,\end{eqnarray*}
\begin{eqnarray*}
H_{r}(A\|B)=-(1-r)^{-1}\ln\left(\sum_{i=1}^{n}a_{i}\frac{a_{i}^{r-1}}{b_{i}^{r-1}}\right),
r > 0, r\neq 1,\end{eqnarray*}
\begin{eqnarray*}
H^{r}(A\|B)=-(1-r)^{-1}\left(\sum_{i=1}^{n}a_{i}\frac{a_{i}^{r-1}}{b_{i}^{r-1}}-1\right),
r > 0, r\neq 1,\end{eqnarray*}
\begin{eqnarray*}_{r}H(A\|B)=-(r-1)^{-1}\left[\left(\sum_{i=1}^{n}a_{i}
\frac{a_{i}^{\frac{1}{r}-1}}{b_{i}^{\frac{1}{r}-1}}\right)^{r}-1\right],
r > 0, r\neq 1,\end{eqnarray*}
\begin{eqnarray*}H(A\|B)=\sum_{i=1}^{n}a_{i}\ln\frac{a_{i}}{b_{i}}\end{eqnarray*}
are the $(r, s)$-relative entropy, the R$\acute{e}$nyi relative
entropy of order $r$, the Tsallis relative entropy of degree $r$,
the relative entropy of type $r$ and the relative
entropy, respectively ([3-5]).

\vskip0.1in

Now, we discuss some elementary properties of the unified
$(r,s)$-relative entropy. First, we point out an important unified
$(r, s)$-directed divergence ${\cal F}_r^s(A\|B)$ which was studied
in [7], note that when $r\neq 1$,
$E^{\frac{s-1}{r-1}}_{r}(A\|B)={\cal F}_r^s(A\|B)$, so by using
Theorem 1 in [7], we can prove the nonnegativity, nonadditivity and
convexity of $E_{r}^{s}(A\|B)$ directly:

\vskip0.1in

(i) $\ $ Let $\Delta_n=\{A=(a_1,a_2, \cdots,a_n): a_i>0,
\sum\limits_{i=1}^{n}a_i=1\}$. If $A,B \in \Delta_n$, then
$E_{r}^{s}(A\|B)\geq 0$, and the equality holds iff $A = B$.

(ii) $\ $ Let $\Delta_m=\{B=(b_1,b_2, \cdots,b_m): b_i>0,
\sum\limits_{i=1}^{m}b_i=1\}$. If $A_1, A_2\in \Delta_n$, $B_1,
B_2\in \Delta_m$, and denote $A* B=(a_{1}b_{1}, \cdots
,a_{1}b_{m},a_{2}b_{1}, \cdots , a_{2}b_{m}, \cdots ,a_{n}b_{m})$,
then
$$E_{r}^{s}(A_1* B_1\|A_2*
B_2)=E_{r}^{s}(A_1\|A_2)+E_{r}^{s}(B_1\|B_2)+(r-1)s
E_{r}^{s}(A_1\|A_2)E_{r}^{s}(B_1\|B_2).$$

(iii) $\ $ If $r=1$ or $r>1, s\geq 1$ or $0< r< 1, s\leq 1$, then
$E_{r}^{s}(A\|B)$ is a convex function of $(A,B)$.

\vskip0.1in

Next, we prove the following:

\vskip0.1in

{\bf Theorem 2.1}. If $r=1$ or $0< r <1, s\geq 0$, then
$$E_{r}^{s}(A\|B)\leq H(A\|B)\leq
E_{2-r}^{s}(A\|B).$$

\noindent {Proof.} That $r=1$ is clear. Let $0<r<1, s=0$. By the
convexity of the function $f(x)=\frac{1}{1-r}\ln x$, we get that
\begin{eqnarray*}H_{r}(A\|B)&=&-(1-r)^{-1}\ln \left(\sum_{i} a_i \frac{{a_{i}}^{r-1}}{{b_i}^{r-1}}\right)
\\&\leq& \sum_i a_i \left[-(1-r)^{-1}\ln \frac{{a_i}^{r-1}}{{b_i}^{r-1}}\right]\\&=&\sum_i a_i\ln \frac{a_i}{b_i}=H(A\|B).\end{eqnarray*}
By a similar way, we get that $H(A\|B)\leq H_{2-r}(A\|B)$. Thus, we
have
\begin{eqnarray}H_r(A\|B)\leq H(A\|B)\leq H_{2-r}(A\|B).\end{eqnarray}

\noindent If $0<r<1$ and $s>0,$ let
$x_0=\sum\limits_{i=1}^{n}a_i\frac{b_{i}^{1-r}}{a_{i}^{1-r}}.$ Then
$\sum\limits_{i=1}^{n}a_i\frac{b_{i}^{1-r}}{a_{i}^{1-r}}\leq
[\sum\limits_{i=1}^{n}(a_i \frac{b_i}{a_i})]^{1-r}=1$. Note that
when $0< x\leq 1$ and $s>0$, we have $\ln x \leq \frac{x^{s}-1}{s}$,
so for any $0<r<1,s>0$, we have $-\frac{x_0^{s}-1}{(1-r)s}\leq
-\frac{1}{1-r}\ln x_0$, thus,
$$-\frac{{\left(\sum\limits_{i=1}^{n}a_i\frac{b_{i}^{1-r}}{a_{i}^{1-r}}\right)}^{s}-1}{(1-r)s}\leq
-\frac{1}{1-r}\ln
\left(\sum\limits_{i=1}^{n}a_i\frac{b_{i}^{1-r}}{a_{i}^{1-r}}\right),$$
that is,
\begin{eqnarray}H_{r}^{s}(A\|B)\leq H_r(A\|B).\end{eqnarray} It follows from
(1) and (2) that $H_{r}^{s}(A\|B)\leq H(A\|B)$. By a similar way, we
can prove $H(A\|B)\leq H_{2-r}^{s}(A\|B).$ Thus, we proved the
theorem.

\vskip0.2in

\noindent {\bf 3. The quantum unified $(r,s)$-relative entropy}

 \vskip0.2in

\noindent Let $H$ be a separable complex Hilbert space and $\rho,
\sigma$ be two states on $H$. Then for any $0\leq r\leq 1$ and any
real number $s$, the quantum unified $(r,s)-$relative entropy is
defined by
\begin{eqnarray*}E_{r}^{s}(\rho\|\sigma)= \left \{
\begin{array}{ll}
H_{r}^{s}(\rho\|\sigma), & {\rm if\ } 0\leq r<1, s\neq 0,\\
H_{r}(\rho\|\sigma), & {\rm if\ } 0\leq r<1, s = 0,\\
H^{r}(\rho\|\sigma), & {\rm if\ } 0\leq r<1, s = 1,\\
_{\frac{1}{r}}H(\rho\|\sigma), & {\rm if\ } 0< r<1, s =\frac{1}{r},\\
H(\rho\|\sigma), & {\rm if\ } r= 1,\end{array} \right.
\end{eqnarray*} where
\begin{eqnarray*}H_{r}^{s}(\rho \| \sigma)=-[(1-r)s]^{-1}[\left(tr (\rho^{r}\sigma^{1-r})\right)^{s}-1], \end{eqnarray*}
\begin{eqnarray*}H_{r}(\rho \| \sigma)=-(1-r)^{-1}\ln \left( tr(\rho^{r}\sigma^{1-r})\right), \end{eqnarray*}
\begin{eqnarray*}H^{r}(\rho \|
\sigma)=-(1-r)^{-1}[tr(\rho^{r}\sigma^{1-r})-1], \end{eqnarray*}
\begin{eqnarray*}_{r}H(\rho \|
\sigma)=-(r-1)^{-1}[\left(tr({\rho^{\frac{1}{r}}\sigma^{1-\frac{1}{r}}})\right)^{r}-1],
\end{eqnarray*}
\begin{eqnarray*}H(\rho\|\sigma)=tr(\rho\ln \rho)- tr(\rho\ln
\sigma)
\end{eqnarray*}
are the quantum $(r, s)$-relative entropy, the quantum
R$\acute{e}$nyi relative entropy of order $r$, the quantum Tsallis
relative entropy, the quantum relative entropy of type $r$ and the
quantum relative entropy ([3-6]), respectively.

\vskip0.2in

We point out that if the state $\sigma$ is invertible, then the
definition of quantum unified $(r,s)-$relative entropy can be
extended to $r>1$. Moreover, we have the following important
equalities:
\begin{eqnarray}H^1_r(\rho\|\sigma)&=&H^r(\rho\|\sigma),\\H^r_{\frac{1}{r}}(\rho\|\sigma)&=&_{r}H(\rho\|\sigma),\end{eqnarray}
(3) and (4) showed that the quantum Tsallis relative entropy and
quantum relative entropy of type $r$ are the particular cases of the
quantum $(r,s)$-relative entropy.

\vskip0.1in

In order to study the properties of quantum unified $(r,s)$-relative
entropy, we need the following lemma.

\vskip0.1in

{\bf Lemma 3.1}.  Let $H$ be a separable complex Hilbert spaces, $A$
and $B$ two positive trace class operators on $H$. Then for any
$\lambda,\mu> 0$, we have $R(\lambda A+\mu B)=R(A+B)$, where $R(A)$
is the range of $A$.

{\bf Proof.} In fact, if $0\leq A\leq B$, that is, $0\leq
A^{\frac{1}{2}}A^{\frac{1}{2}}\leq B^{\frac{1}{2}}B^{\frac{1}{2}}$,
then it follows from Theorem 1 in [8] that
$R({A}^{\frac{1}{2}})\subseteq R({B}^{\frac{1}{2}})$. Note that
$R({A}^{\frac{1}{2}})=R({A})$, $R({B}^{\frac{1}{2}})=R(B)$, so
$R(A)\subseteq R(B)$. Thus, we have $R(A+B)\subseteq
R(A+(1+\alpha)B)$ for any $\alpha > 0$. On the other hand, take $n$
such that $0\leq \frac{A+(1+\alpha)B}{n}\leq A+B$, then
$R(A+(1+\alpha)B)=R(\frac{A+(1+\alpha)B}{n})\subseteq R(A+B).$
 Thus, $R(A+B)=R(A+(1+\alpha)B)$, i.e., $R(A+B)=R(A+\beta B)$ for any $\beta>1.$
 Replace $B$ with $\frac{1}{\beta}B$, we have $R(A+\frac{1}{\beta}B)=R(A+B)$ for any $\beta> 1$.
 Hence, $R(A+\mu B)= R(A+B)$ for any $\mu> 0$. Furthermore, $R(\lambda A+\mu B)=R(A+\mu
 B)=R(A+B)$ for any $\lambda >0$ and $\mu> 0$.

{\bf Theorem 3.1}. Let $H$, $H_1$ and $H_2$ be separable complex
Hilbert spaces.

(I) $ \ $ If $\rho$ and $\sigma$ are two states on $H$, then
$E_{r}^{s}(\rho\|\sigma)\geq 0$. Furthermore, when $0< r\leq 1$,
$E_{r}^{s}(\rho\|\sigma)=0$ iff $\rho=\sigma$; when $r=0$,
$E_{r}^{s}(\rho\|\sigma)=0$ iff $Ker (\rho)\subseteq Ker (\sigma)$.

(II) $ \ $  If $\rho_j$ and $\sigma_j$ are states on $H$,
$\lambda_i>0$, $j=1, 2, \cdots, n$, and $\sum_{j=1}^n\lambda_j=1$,
then when $r=1$ or $0\leq r< 1$ and $s\leq 1$, we have
$$E_{r}^{s}(\sum_{j}\lambda_{j}\rho_{j}\|\sum_{j}\lambda_{j}\sigma_{j})\leq
\sum_{j}\lambda_{j}E_{r}^{s}(\rho_{j}\|\sigma_{j}).$$

(III) $ \ $ If $\rho$ and $\sigma$ are two states, $U$ is a unitary
operator on $H$, then
$$E_{r}^{s}(U\rho U^{*}\|U\sigma U^{*})=E_{r}^{s}(\rho\| \sigma).$$

(IV) $ \ $ If $\rho_1$ and $\sigma_1$ are two states on $H_1$,
$\rho_2$ and $\sigma_2$ are two states on $H_2$, then
$$E_{r}^{s}(\rho_{1}\otimes \rho_{2}\|\sigma_{1}\otimes
\sigma_{2})=E_{r}^{s}(\rho_{1}\|\sigma_{1})+E_{r}^{s}(\rho_{2}\|\sigma_{2})
+(r-1)sE_{r}^{s}(\rho_{1}\|\sigma_{1})E_{r}^{s}(\rho_{2}\|\sigma_{2}).$$

 {\bf Proof.} For $r=1$, the conclusion had been proved (see [3], [9-12]). Note that (3) and (4),
 we only need to prove the cases of $E^{s}_{r}(\rho\|\sigma)=H^{s}_{r}(\rho\|\sigma)$ and
 $E^{s}_{r}(\rho\|\sigma)=H_{r}(\rho\|\sigma)$.

 (I) $ \ $ Note that when $0\leq r < 1$ and $s\neq 0$, $h(x)=\frac{1-x^{s}}{(1-r)s}$ and
$g(x)=\frac{\ln x}{r-1}$ are monotone decreasing, so it is
sufficient to prove that $0\leq tr(\rho^r\sigma^{1-r}) \leq 1$.

Let $\rho=0P_0+\sum\limits_{i}\lambda_i P_i$ and
$\sigma=0Q_0+\sum\limits_{j} \mu_j Q_j$ be the spectral
decompositions of states $\rho$ and $\sigma$, where $i, j\in {\bf
N}=\{1,2,\cdots\}$, $P_i$ and $Q_j$ are the one dimension projection
operators, $P_0$ and $Q_0$ are the projections on the kernel spaces
of $\rho$ and $\sigma$, respectively, and $\lambda_i>0, \mu_j>0$.
Then $P_iP_0=0$, $Q_jQ_0=0$, $P_iP_j=Q_iQ_j=0$ if $i\neq j$,
$P_0+\sum\limits_{i}P_i=Q_0+\sum\limits_{j}Q_j=I$ and
$\sum\limits_{i}\lambda_i=\sum\limits_{j}\mu_j=1$. So, we have
$tr(P_0Q_j)+\sum\limits_{i} tr(P_iQ_j)=tr(Q_j)=1$ and
$tr(Q_0P_i)+\sum\limits_{j} tr(P_iQ_j)=tr(P_i)=1$. Thus, when $0\leq
r<1$,
\begin{eqnarray*}tr(\rho^r\sigma^{1-r})&=&\sum_{i}\sum_{j}\lambda_i^r
\mu_j^{1-r}tr (P_i Q_j)\\&=&\sum_{i}\sum_{j}\lambda_i^r
\mu_j^{1-r}tr (Q_j P_i Q_j)\\&\geq& 0.
\end{eqnarray*}

When $r=0,$ \begin{eqnarray*}tr(\rho^0\sigma^{1-0})&=&\sum_{ij}\mu_j
tr(P_i Q_j)\\&=&\sum_{j}\mu_j tr(\sum_{i}P_i Q_j)\\&\leq
&\sum_{j}\mu_j=1,\end{eqnarray*} and with equality iff for any $j,
\sum\limits_i tr(P_i Q_j)=1$ iff for any $j, tr (P_0 Q_j)=0$ iff $
P_0\leq Q_0$ iff $Ker (\rho)\subseteq Ker (\sigma)$.

When $0<r<1$, note that $\sum_{i}tr (P_i Q_j)+tr (P_0 Q_j)=1,$ by
the concavity of $f(x)=x^r$,  we have
\begin{eqnarray}tr(\rho^r\sigma^{1-r})&=&\sum_{i}\sum_{j}\lambda_i^r
\mu_j^{1-r}tr (P_i Q_j)\\&=&\sum_{j} \mu_j [\sum_{i}
(\frac{\lambda_i}{\mu_j})^{r}tr (P_i Q_j)+(\frac{0}{\mu_j})^{r}tr
(P_0 Q_j)]\\&\leq &\sum_{j} \mu_j [\sum_i\frac{\lambda_i}{\mu_j}tr
(P_iQ_j)+\frac{0}{\mu_j}tr (P_0 Q_j)]^r\\&\leq& (\sum_{j} \mu_j
{\sum_{i}\frac{\lambda_i}{\mu_j}tr
(P_iQ_j)})^r\\&=&(\sum_{i}\lambda_i\sum_{j} tr (P_iQ_j))^r\\&\leq
&1.
\end{eqnarray}

Thus, we proved that when $0\leq r<1$, $0\leq tr(\rho^r\sigma^{1-r})
\leq 1$, and when $r=0$, $tr(\rho^0\sigma^{1-0})=1$ iff $Ker
(\rho)\subseteq Ker (\sigma)$. Note that when $0<r<1$,
$E_{r}^{s}(\rho\|\sigma)=0$ iff $tr(\rho^r\sigma^{1-r})=1$, so, we
only need to prove that if $0<r<1$ and $tr(\rho^r\sigma^{1-r})=1$,
then $\rho=\sigma$.

First, if $tr(\rho^r\sigma^{1-r})=1$, it follows from (9) and (10)
that for each $i\in\bf N$, $\sum\limits_{j} tr (P_iQ_j)=1$, so $tr
(P_iQ_0)=0=tr (P_iQ_0P_i)$, it is easily to know that $
P_iQ_0P_i=0$, so for each $i\in\bf N$, $ P_iQ_0=0$, thus we have
$Q_0\leq P_0$. Moreover, if $tr(\rho^r\sigma^{1-r})=1$, then (7)
takes equality, we get that (i) or (ii) as follows:

(i) $ \ $ For each given $j$, there exists a $i_j\in\bf N$ such that
$tr(P_{i_j}Q_j)=1$ and $tr(P_iQ_j)=0$ for all $i\neq i_j$.

(ii) $ \ $ For each $j$, we have
$\frac{\lambda_1}{\mu_j}=\frac{\lambda_{2}}{\mu_j}=\cdots$ and
$\sum\limits_{i}tr (P_i Q_j)=1$.

If (i) is satisfied, then for each $j$, we have $P_0Q_j=0$, so
$P_0\leq Q_0$, combining this and $Q_0\leq P_0$ proved before, we
get $Q_0=P_0$. Moreover, note that $P_{i_j}$ and $Q_j$ are both one
dimensional projections and $tr(P_{i_j}Q_j)=1$, so, it is easy to
know that $P_{i_j}=Q_j$. It also follows  from
$tr(\rho^r\sigma^{1-r})=1$ that $\frac{\lambda_{i_j}}{\mu_j}=1$,
thus, we can prove that $\rho=\sigma$.

If (ii) is satisfied, then for each $j$, we have
$\frac{\lambda_1}{\mu_j}=\frac{\lambda_{2}}{\mu_j}=\cdots$ and
$\sum_{i}tr (P_i Q_j)=1$, so we have $\lambda_1=\lambda_{2}=\cdots$
and for each $j$, $tr(P_0Q_j)=0$, so we can prove that $P_0\leq
Q_0$, thus, $P_0=Q_0$. Moreover, it follows from
$\frac{\lambda_i}{\mu_j}$ is a constant, $\sum_{i}tr (P_i Q_j)=1$
and (5)-(10) that $\mu_1=\mu_2=\cdots=\lambda_1=\lambda_{2}=\cdots$,
thus, we have $\rho=\sigma$, (I) is proved.

\vskip0.1in

(II) $ \ $  Let $\rho$ and $\sigma$ be two states on $H$ and
$f(\rho,\sigma)=tr({\rho}^{r}{\sigma}^{1-r})$. If $0< r< 1$, then it
follows from [13, Corollary 1.1] that
$f(\rho,\sigma)=tr({\rho}^{r}{\sigma}^{1-r})$ is a joint concave
functional with respect to the states $\rho$ and $\sigma$, that is,
for any states $\rho_1$, $\rho_2$, $\sigma_1$ and $\sigma_2$, when
$0< \lambda<1$, we have
\begin{eqnarray} f(\lambda \rho_{1}+(1-\lambda)\rho_{2},\lambda
\sigma_{1}+(1-\lambda)\sigma_{2})\geq \lambda
f(\rho_{1},\sigma_{1})+(1-\lambda)f(\rho_{2},\sigma_{2}).
\end{eqnarray}

If $r=0$, let $P_1$, $P_2$ and $P$ be the projection operators on
$R(\rho_1)$, $R(\rho_2)$ and $R(\lambda\rho_1+(1-\lambda)\rho_2)$,
respectively, then $\rho_1^0=P_1$, $\rho_2^0=P_2$,
$(\lambda\rho_1+(1-\lambda)\rho_2)^0=P$. It follows from Lemma 3.1
that $P\geq P_1$ and $P\geq P_2.$ Therefore, we have
\begin{eqnarray*} &\
&tr((\lambda\rho_1+(1-\lambda)\rho_2)^0(\lambda
\sigma_1+(1-\lambda)\sigma_2)^1)\\&=&tr(P(\lambda
\sigma_1+(1-\lambda)\sigma_2))\\&=&\lambda
tr(P\sigma_1)+(1-\lambda)tr(P\sigma_2)\\&\geq&\lambda
tr(P_1\sigma_1)+(1-\lambda)tr(P_2\sigma_2)\\&=&\lambda
tr(P_1\sigma_1)+(1-\lambda)tr(P_2\sigma_2)\\&=&\lambda
tr((\rho_1)^0\sigma_1)+(1-\lambda)tr((\rho_2)^0\sigma_2).\end{eqnarray*}

This shows that the inequality (11) also holds when $r=0$.

\vskip0.1in

If $0\leq r< 1, s=0$, by the monotone decreasing property and
convexity of the function $g(x)=\frac{\ln x}{r-1}$, we have
\begin{eqnarray*}&\quad&H_r(\lambda \rho_1+(1-\lambda)\rho_2\|\lambda \sigma_1+(1-\lambda)\sigma_2
)\\&=&\frac{1 }{r-1}\ln(f(\lambda
\rho_{1}+(1-\lambda)\rho_{2},\lambda
\sigma_{1}+(1-\lambda)\sigma_{2}))\\&\leq&\frac{1}{r-1}\ln(\lambda
f(\rho_{1},\sigma_{1})+(1-\lambda)f(\rho_{2},\sigma_{2}))\\&\leq&\lambda\frac{1
}{r-1}\ln(f(\rho_{1},\sigma_{1})+(1-\lambda)\frac{1
}{r-1}\ln(f(\rho_{2},\sigma_{2}))\\&=&\lambda
H_r(\rho_1\|\sigma_1)+(1-\lambda)H_r(\rho_2\|\sigma_2).
\end{eqnarray*}

If $0\leq r< 1, s\neq 0$ and $s\leq 1$, then
$h(x)=\frac{1-x^{s}}{(1-r)s}$ is also a monotone decreasing convex
function, so
\begin{eqnarray*}&\quad&H_{r}^{s}(\lambda \rho_{1}+(1-\lambda)\rho_{2}\|\lambda \sigma_{1}+(1-\lambda)\sigma_{2})
\\&=&[(1-r)s]^{-1}[1-f^{s}(\lambda \rho_{1}+(1-\lambda)\rho_{2},\lambda \sigma_{1}+(1-\lambda)\sigma_{2})]
\\&\leq &[(1-r)s]^{-1}[1-(\lambda f^{s}( \rho_{1},\sigma_{1})+(1-\lambda)f^{s}(\rho_{2},\sigma_{2}))]\\&=&\lambda
H_r^s(\rho_1\|\sigma_1)+(1-\lambda)H_r^s(\rho_2\|\sigma_2).\end{eqnarray*}
Thus, (II) is proved. (III) and (IV) can be proved easily, we omit
them.

\vskip0.1in

In order to study the other properties of quantum unified
$(r,s)$-relative entropy, we need the following:

\vskip0.1in

Let $H_1$ and $H_2$ be two separable complex Hilbert spaces and
$H_1\otimes H_2$ their tensor product. The set of all trace class
operators on $H_1\otimes H_2$ is denoted by $T(H_1\otimes H_2)$, the
set of all trace class positive operators on $H_1\otimes H_2$ is
denoted by $T_{+}(H_1\otimes H_2)$. If $A\in T_{+}(H_1\otimes H_2)$,
by the following form, we can define a trace class positive operator
$A_1$ on $H_1$: $$(x,A_1 y)=\sum_i (x\otimes e_i, A(y\otimes e_i)),
$$ where $x, y\in H_1$, $\{e_i\}$ is an orthonormal basis of $H_2$.
We call $A_1$ to be the {\it partial trace} of $A$ on $H_1$ and
denoted by $A_1=tr_2 A$. Similarly, we can define the {\it partial
trace} $A_2$ of $A$ on $H_2$. Note that when $A$ is a state, $A_1$
and $A_2$ are also states.

\vskip0.1in

It follows from Theorem 3.1(III), Theorem 3.1(IV) and the methods in
the proof of [3, Theorem 11.17], we have

\vskip0.1in

{\bf Lemma 3.2}. Let $H_1, H_2$ be two finite dimensional complex
Hilbert space. If $r=1$ or $0\leq r< 1$ and $s\leq 1$, then for any
states $\rho$ and $\sigma$ on $H_1\otimes H_2$,
$$E_r^s(\rho_1||\sigma_1)\leq E_r^s(\rho||\sigma),$$ where $\rho_1$ and
$\sigma_1$ are the partial traces of $\rho$ and $\sigma$ on $H_1$,
respectively.

\vskip0.1in

{\bf Lemma 3.3}. Let $H$ be a finite dimensional complex Hilbert
space, $\Phi$ a trace-preserving completely positive map of $T(H)$
into itself. If $r=1$ or $0\leq r< 1$ and $s\leq 1$, then for any
states $\rho$ and $\sigma$ on $H$,
$$E_r^s(\Phi (\rho)||\Phi (\sigma))\leq E_r^s(\rho||\sigma).$$

{\bf Proof.} Taking a finite dimensional complex Hilbert space $H_0$
such that the dimension of $H_0$ is bigger than 1. Then it follows
from ([9,11-12]) that there are a unitary operator $U$ on $H\otimes
H_0$ and a projection operator $P$ on $H_0$ such that for any state
$\rho$ on $H$, we have $$\Phi(\rho)=tr_2 (U(\rho\otimes P)U^*),$$
thus, it follows from Lemma 3.2 that
$$E_r^s(\Phi (\rho)||\Phi (\sigma))\leq E_r^s(U(\rho\otimes P)U^*||U(\sigma\otimes P)U^*)=E_r^s(\rho\otimes P||\sigma\otimes P)=E_r^s(\rho||\sigma).$$

\vskip0.1in

{\bf Lemma 3.4} ([11]). Let $H$ be a separable complex Hilbert
space, $\Phi$ a trace-preserving completely positive map of $T(H)$
into itself, and $\{P_n\}$ a family of finite-dimensional
projections such that $P_m\leq P_n$ for $m\leq n$ and
$P_n\rightarrow I$ strongly when $n\rightarrow \infty$. Then there
is a family $\{\Phi_n\}$ of completely positive maps such that
$\{\Phi_n\}$ is trace-preserving on $P_n (H)$ and $\Phi_n
(A)\rightarrow \Phi (A)$ uniformly for each $A\in T_{+}(H)$.

\vskip0.1in

{\bf Theorem 3.2}. Let $H$ be a separable complex Hilbert space and
$\Phi$ a trace-preserving completely positive map of $T(H)$ into
itself. If $r=1$ or $0\leq r< 1$ and $s\leq 1$, then for any state
$\rho$ and $\sigma$, we have
\begin{eqnarray*}E_r^s(\Phi (\rho)\|\Phi (\sigma))\leq E_r^s(\rho\|\sigma).\end{eqnarray*}

{\bf Proof.} Let $P_n$ and $\Phi_n$ satisfy the conditions of Lemma
3.4. Then $\Phi_n (\rho)\rightarrow \Phi (\rho)$ uniformly for each
 state $\rho$. Since function $x^ry^{1-r}$ is continuous, we have $(\Phi_n
(\rho))^r(\Phi_n (\sigma))^{1-r}\rightarrow (\Phi (\rho))^r(\Phi
(\sigma))^{1-r}$ uniformly, hence $tr((\Phi_n (\rho))^r(\Phi_n
(\sigma))^{1-r})\rightarrow tr((\Phi (\rho))^r(\Phi
(\sigma))^{1-r}).$ This shows that
\begin{eqnarray*}E_r^s(\Phi (\rho)\|\Phi (\sigma))=\lim_{n\rightarrow \infty}E_r^s(\Phi_n
(\rho)\|\Phi_n (\sigma)).\end{eqnarray*} Let $\rho_n=\frac{P_n \rho
P_n}{tr(\rho P_n)}, \sigma_n=\frac{P_n \sigma P_n}{tr(\sigma P_n)}$.
By the proof of Lemma 4 in [10], $P_n\rho P_n\rightarrow \rho$ and
$P_n \sigma P_n\rightarrow \sigma$ uniformly. Hence $tr(P_n\rho
P_n)\rightarrow tr(\rho)=1$ and $tr(P_n\sigma P_n)\rightarrow
tr(\sigma)=1.$ Therefore
\begin{eqnarray*}\lim_{n\rightarrow
\infty}(\Phi_n(\rho))^r(\Phi_n(\sigma))^{1-r}&=&
\lim_{n\rightarrow \infty}(\Phi_n(P_n\rho P_n))^r(\Phi_n(P_n\sigma P_n))^{1-r}
\\&=&\lim_{n\rightarrow \infty}\frac{(\Phi_n(P_n\rho P_n))^r(\Phi_n(P_n\sigma P_n))^{1-r}}
{(tr(P_n\rho P_n))^r(tr(P_n\sigma P_n))^{1-r}}\\&=&\lim_{n\rightarrow \infty}(\Phi_n(\rho_n))^r(\Phi_n(\sigma_n))^{1-r}.
\end{eqnarray*}
Hence we get that $\lim\limits_{n\rightarrow
\infty}E_r^s(\Phi_n(\rho)\|\Phi_n(\sigma))=\lim\limits_{n\rightarrow
\infty}E_r^s(\Phi_n(\rho_n)\|\Phi_n(\sigma_n)).$
 By Lemma 3.2, $E_r^s(\Phi_n (\rho_n)\|\Phi_n (\sigma_n))\leq
E_r^s(\rho_n\|\sigma_n)$. Again $\rho_n\rightarrow \rho,
\sigma_n\rightarrow \sigma$ uniformly, we get that
$\lim\limits_{n\rightarrow
\infty}E_r^s(\rho_n\|\sigma_n)=E_r^s(\rho\|\sigma).$ Therefore
$E_r^s(\Phi (\rho)\|\Phi (\sigma))\leq E_r^s(\rho\|\sigma).$ That
completes the proof.

\vskip0.1in

{\bf Theorem 3.3 (Monotonicity).} Let $H_1, H_2$ be separable
complex Hilbert space, $H=H_1\otimes H_2$. If $r=1$ or $0\leq r<1$
and $s\leq 1$, then for any state $\rho$ and $\sigma$ on $H$,
$$E_r^s(\rho_1\|\sigma_1)\leq E_r^s(\rho\|\sigma),$$ where $\rho_1$ and
$\sigma_1$ are the partial traces of $\rho$ and $\sigma$ on $H_1$,
respectively.

{\bf Proof.} Since $H_2$ is a separable complex Hilbert space, so
there is a sequence of $\{P_n\}$ of finite-dimensional projection
operators on $H_2$ such that $P_m\leq P_n$ for $m\leq n$ and
$P_n\rightarrow I$ strongly when $n\rightarrow \infty$. Let
$H_2^n=P_n (H_2), H^n=H_1 \otimes H_2^n$. It follows from the proof
of Lemma 4 in [10] again that $$\rho_n=\frac{(I\otimes
P_n)\rho(I\otimes P_n)}{tr(\rho(I\otimes P_n))}\rightarrow \rho,$$
$$\sigma_n=\frac{(I\otimes P_n)\sigma(I\otimes P_n)}{tr(\sigma(I\otimes P_n))}\rightarrow \sigma,$$
$$\rho_{1n}=tr_2 \rho_n\rightarrow \rho_1,$$
$$\sigma_{1n}=tr_2 \sigma_n\rightarrow \sigma_1$$ uniformly.
Hence $E_r^s(\rho_{1n}\|\sigma_{1n})\rightarrow
E_r^s(\rho_1\|\sigma_1),$ and
$E_r^s(\rho_{n}\|\sigma_{n})\rightarrow E_r^s(\rho\|\sigma).$

Define $\Phi:B(H^n)\rightarrow B(H_1)\otimes \{\lambda I_2^n\}$ by
$\Phi(\rho)=(tr_2\rho)\otimes C_{2n},$ where $I_2^n$ is the identity
operator on $H_2^n$) and $C_{2n}=(dim H_2^n)^{-1}I^n_2.$ Then
$$E_r^s(\Phi (\rho_n)\|\Phi(\sigma_n))=E_r^s(\rho_{1n}\otimes
C_{2n}\|\sigma_{1n}\otimes C_{2n})=E_r^s(\rho_{1n}\|\sigma_{1n}).$$
It is obvious that $\Phi$ is a trace-preserving completely positive
map from $B(H^n)$ into itself. By Theorem 3.2, $E_r^s(\Phi
(\rho_n)\|\Phi(\sigma_n))\leq E_r^s(\rho_n\|\sigma_n)$, so
$$E_r^s(\rho_{1n}\|\sigma_{1n})\leq E_r^s(\rho_n\|\sigma_n).$$
Note that $E_r^s(\rho_{1n}\|\sigma_{1n})\rightarrow
E_r^s(\rho_1\|\sigma_1),$ $E_r^s(\rho_{n}\|\sigma_{n})\rightarrow
E_r^s(\rho\|\sigma),$ thus we have $$E_r^s(\rho_1\|\sigma_1)\leq
E_r^s(\rho\|\sigma)$$ and the theorem is proved.

\vskip0.1in

{\bf Theorem 3.4}. Let $H$ be a separable complex Hilbert space,
$\rho$ and $\sigma$ two states on $H$ and $\sigma$ invertible.
 Then for $r=1$ or $0\leq r < 1, s\geq 0,$ we have
 \begin{eqnarray}E_{r}^{s}(\rho\|\sigma)\leq H(\rho\|\sigma)\leq E_{2-r}^{s}(\rho\|\sigma).\end{eqnarray}

{\bf Proof.} That $r=1$ is clear. If $0\leq r< 1, s=0$, we need to
prove that
\begin{eqnarray} H_{r}(\rho\|\sigma)\leq H(\rho\|\sigma)\leq
H_{2-r}(\rho\|\sigma). \end{eqnarray}

Let $\rho=0P_0+\sum\limits_{i}\lambda_i P_i$ and
$\sigma=\sum\limits_{j} \mu_j Q_j$ be the spectral decompositions of
$\rho$ and $\sigma$, where $P_i$ and $Q_j$ be the one dimension
projection operators, $P_0$ be the projection operator on the kernel
space of $\rho$, and $P_iP_0=0$, $\lambda_i>0, \mu_j>0$ when $i,
j\in \bf N$, and $P_iP_j=Q_iQ_j=0$ if $i\neq j$,
$\sum\limits_{i}\lambda_i=\sum\limits_{j}\mu_j=1$,
$P_0+\sum\limits_{i}P_i=\sum\limits_{j}Q_j=I$. Then
\begin{eqnarray*}H_{2-r}(\rho\|\sigma)&=&-\frac{\ln tr(\rho^{2-r}\sigma^{r-1})}{r-1}
\\&=&-\frac{1}{r-1}\ln \sum_{ij}\lambda_{i}^{2-r}\mu_{j}^{r-1}tr(P_{i}Q_{j})
\\&=&-\frac{1}{r-1}\ln \sum_{ij}\lambda_{i}tr(P_{i}Q_{j})(\frac{\lambda_{i}}{\mu_{j}})^{1-r}\end{eqnarray*}

Let $g(x)=-\frac{1}{r-1}\ln x,\
\alpha_{ij}=\lambda_{i}tr(P_{i}Q_{j})$ and $
x_{ij}=(\frac{\lambda_{i}}{\mu_{j}})^{1-r}.$ Then
$\sum\limits_{ij}\alpha_{ij}=\sum\limits_{ij}\lambda_{i}tr(P_{i}Q_{j})=\sum\limits_{j}tr(\rho
Q_{j})=tr(\rho)=1.$ By the concavity of the function
$g(x)=-\frac{1}{r-1}\ln x,$ we have
\begin{eqnarray*}H_{2-r}(\rho\|\sigma)&=&g(\sum_{ij}\alpha_{ij}x_{ij})\\&\geq&\sum_{ij}\alpha_{ij}g(x_{ij})
\\&=&\sum_{ij}\lambda_{i}tr(P_{i}Q_{j})\left(-\frac{1}{r-1}\ln (\frac{\lambda_{i}}{\mu_{j}})^{1-r}\right)
\\&=&\sum_{ij}\lambda_{i}tr(P_{i}Q_{j})(\ln \lambda_{i}-\ln \mu_{j})\\&=&H(\rho\|\sigma).\end{eqnarray*}
The left-hand side inequality of (13) is proven by a similar way.

If $0\leq r< 1, s>0$, we need to prove that
\begin{eqnarray}H_{r}^{s}(\rho\|\sigma)\leq H(\rho\|\sigma)\leq
H_{2-r}^{s}(\rho\|\sigma).\end{eqnarray}
 Let $tr(\rho^{r}\sigma^{1-r})=x_0.$ Since
$$\ln x \leq \frac{x^{s}-1}{s},$$ for any $x> 0,
s>0,$ so, for any $0\leq r<1,s>0$, we have
$$-\frac{x_0^{s}-1}{[(r-1)s]}\geq -(r-1)^{-1}\ln x_0.$$
That is, $$H_{2-r}^{s}(\rho\|\sigma)\geq H_{2-r}(\rho\|\sigma).$$
Combining this with (13), we have $H_{2-r}^{s}(\rho\|\sigma)\geq
H(\rho\|\sigma)$. Similarly, the left-hand side inequality of (14)
can be proven.

\vskip0.1in

Note that when $0\leq r< 1, s=1$, the inequalities (12) degenerate
into convexity inequalities for estimating free energy and relative
entropy given by Ruskai and Stillinger in [14].

\vskip0.1in

 {\bf Theorem 3.5}. Let $\rho$ and $\sigma$ be two states on the separable complex Hilbert space ${\mathcal{H}}$. We have

(1) $ \ $ If $\rho$ is an invertible state, then
$E_0^s(\rho\|\sigma)=0$.

(2) $ \ $  If $s\geq 0$, then $E_r^s(\rho\|\sigma)$ is monotone
increasing with respect to $r\in [0, 1]$; if $s< 0$, then
$E_r^s(\rho\|\sigma)$ is monotone decreasing with respect to $r\in
[0, 1]$.

(3) $ \ $ For each $0\leq r\leq 1$, $E_r^s(\rho\|\sigma)$ is
monotone decreasing with respect to $s$.

(4) $ \ $  For each $0\leq r\leq 1$, $E_r^s(\rho\|\sigma)$ is a
convex function of $s$.

{\bf Proof.} (1) $ \ $ If $\rho$ is invertible, we have $\rho^0=I$,
so $E_0^s(\rho\|\sigma)=0$.

(2) $ \ $ It follows from Theorem 3.4 that
$E_{r}^{s}(\rho\|\sigma)\leq
H(\rho\|\sigma)=E_{1}^{s}(\rho\|\sigma)$, so it is sufficient to
prove the conclusion for $0\leq r<1$ and any
$s$.

Let $\rho=0P_0+\sum\limits_{i}\lambda_i P_i$ and
$\sigma=0Q_0+\sum\limits_{j} \mu_j Q_j$ be the spectral
decompositions of $\rho$ and $\sigma$, where $P_i$ and $Q_j$ are the
one dimension projection operators, $P_0$ and $Q_0$ are the
projection operators on the zero spaces of $\rho$ and $\sigma$
respectively, and for all $i, j\in \bf N$, $\lambda_i>0, \mu_j> 0$.
Then $\sum\limits_{i}\lambda_i=\sum\limits_{j}\mu_j=1$,
$P_0+\sum\limits_{i}P_i=Q_0+\sum\limits_{j}Q_j=I$.

Let $f(x)=x\ln x$, $\alpha_{ij}=\lambda_{i}tr (P_{i}Q_{j})$, $
x_{ij}=(\frac{\mu_{j}}{\lambda_{i}})^{1-r}$. Then
$\sum\limits_{i}[\sum\limits_{j}\alpha_{ij}+\lambda_{i}tr
(P_{i}Q_{0})]= \sum\limits_{i}\lambda_{i}tr
(P_{i}(\sum\limits_{i}Q_{j}+Q_0))=tr(\rho)=1.$ Because $f(x)$ is a
convex function, we have $\sum\limits_{ij}\alpha_{ij}f(x_{ij})\geq
f(\sum\limits_{ij}\alpha_{ij}x_{ij}).$ Therefore,
$$\sum\limits_{ij}\lambda_{i}tr
(P_{i}Q_{j})(\frac{\mu_{j}}{\lambda_{i}})^{1-r}\ln
(\frac{\mu_{j}}{\lambda_{i}})^{1-r}\geq
\sum\limits_{ij}\lambda_{i}tr
(P_{i}Q_{j})(\frac{\mu_{j}}{\lambda_{i}})^{1-r}\ln
\sum\limits_{ij}\lambda_{i}tr(P_{i}Q_{j})
(\frac{\mu_{j}}{\lambda_{i}})^{1-r},$$ that is,
\begin{eqnarray}-(1-r)tr(\rho^{r}(\ln\rho-\ln\sigma)\sigma^{1-r})\geq tr(\rho^{r}\sigma^{1-r})\ln
tr(\rho^{r}\sigma^{1-r}).
\end{eqnarray}

(i) $ \ $ If $s=0$, then
$E_r^0(\rho\|\sigma)=H_r(\rho\|\sigma)=-(1-r)^{-1}\ln (
tr(\rho^{r}\sigma^{1-r}))$. Note that $tr(\rho^r \sigma^{1-r})=0$
iff for any $i,j\in {\bf N}, P_i Q_j=0.$ Hence, if for some $0\leq
r_0<1$ such that $tr(\rho^{r_0} \sigma^{1-{r_0}})=0$, then it is
easily to see that for any $0\leq r<1$, $tr(\rho^{r}
\sigma^{1-{r}})=0$, so for any $0\leq r<1$,
$E_r^0(\rho\|\sigma)=H_r(\rho\|\sigma)=+\infty$, thus, the
conclusion is also true in this case. If for each $0\leq r<1$,
$tr(\rho^r \sigma^{1-r})>0$, then
\begin{eqnarray*}\frac{dH_{r}(\rho\|\sigma)}{dr}&=&\frac{H_{r}(\rho\|\sigma)}{1-r}-\frac{tr(\rho^{r}(\ln \rho-\ln \sigma)\sigma^{1-r})}
{(1-r)tr(\rho^{r}\sigma^{1-r})}\\&=&\frac{-\ln tr(
\rho^{r}\sigma^{1-r})}{(1-r)^{2}}-\frac{tr(\rho^{r}(\ln \rho-\ln
\sigma)\sigma^{1-r})}
{(1-r)tr(\rho^{r}\sigma^{1-r})}.\end{eqnarray*} By (15), we know
that $-tr(\rho^{r}(\ln \rho-\ln \sigma)\sigma^{1-r})\geq
\frac{1}{1-r}{tr(\rho^{r}\sigma^{1-r})\ln
tr(\rho^{r}\sigma^{1-r})},$ so $$\frac{dH_{r}(\rho\|\sigma)}{dr}\geq
\frac{-\ln tr(\rho^{r}\sigma^{1-r})}{(1-r)^{2}}+\frac{\ln
tr(\rho^{r}\sigma^{1-r})}{(1-r)^{2}}=0.$$ This conclusion is proved
when $s=0$.

(ii) $ \ $ If $s\neq 0$, and for some $0\leq r_0<1$, $tr(\rho^{r_0}
\sigma^{1-{r_0}})=0$, then for any $0\leq r<1$, $tr(\rho^{r}
\sigma^{1-{r}})=0$, so for any $0\leq r<1$,
$E_r^s(\rho\|\sigma)=H_r^s(\rho\|\sigma)=\frac{1}{(1-r)s}$, thus,
the conclusion is true in this case. If for each $0\leq r<1$,
$tr(\rho^r \sigma^{1-r})>0$,  then
$E_r^s(\rho\|\sigma)=H_r^s(\rho\|\sigma)=-[(1-r)s]^{-1}[
(tr(\rho^{r}\sigma^{1-r}))^s-1]$, so by (15) again that
\begin{eqnarray*}&\
&\frac{dH_{r}^{s}(\rho\|\sigma)}{dr}\\&=&\frac{H_{r}^{s}(\rho\|\sigma)}{1-r}
-\frac{(tr (\rho^{r}\sigma^{1-r}))^{s-1}tr\rho^{r}(\ln \rho-\ln
\sigma)\sigma^{1-r}}{1-r}\\&=&\frac{1}{(1-r)^{2}}\left[\frac{1-(tr
(\rho^{r}\sigma^{1-r}))^{s}}{s}-(1-r)(tr
(\rho^{r}\sigma^{1-r}))^{s-1}tr(\rho^{r}(\ln \rho-\ln
\sigma)\sigma^{1-r})\right]\\&\geq&\frac{1}{(1-r)^{2}}\left[\frac{1-(tr
(\rho^{r}\sigma^{1-r}))^{s}}{s}+\frac{s(tr (\rho^{r}\sigma^{1-r})\ln
(tr \rho^{r}\sigma^{1-r}))(tr
(\rho^{r}\sigma^{1-r}))^{s-1}}{s}\right]\\&=&\frac{1}{(1-r)^{2}}\left[\frac{1-(tr
(\rho^{r}\sigma^{1-r}))^{s}+s(tr (\rho^{r}\sigma^{1-r}))^{s}\ln tr
(\rho^{r}\sigma^{1-r})}{s}\right].\end{eqnarray*}

Let $tr(\rho^{r}\sigma^{1-r})=x,\ f(x)=\frac{sx^{s}\ln
x-x^{s}+1}{s}.$ Then $0<x\leq 1,\ f'(x)=sx^{s-1}\ln x.$ Note that
If $s> 0$, then $f'(x)\leq 0,\ f(x)\geq f(1)=0.$ Thus,
$\frac{dH^{s}_{r}(\rho\|\sigma)}{dr}\geq 0.$
Similarly, $\frac{dH^{s}_{r}(\rho\|\sigma)}{dr}\leq 0$ if $s< 0$.

The conclusion is
proved finally.

(3) $ \ $ If $r=1$, then $E_r^s(\rho\|\sigma)=H(\rho\|\sigma)$ is a constant, so the conclusion is true in this case.

If
for some $0\leq r_0<1$, $tr(\rho^{r_0} \sigma^{1-{r_0}})=0$, then
for any $0\leq r<1$, $tr(\rho^{r} \sigma^{1-{r}})=0$, thus,
$E_r^0(\rho\|\sigma)=+\infty
>\frac{1}{(1-r)s}=E_r^s(\rho\|\sigma)$ for any $s$.

If for each
$0\leq r<1$, $tr(\rho^r \sigma^{1-r})>0$, by the inequality $$\ln x\leq \frac{x^s-1}{s}$$ for any $x> 0, s>0$, we can prove that for any $s>0$ and $0\leq r<1$,
$E_r^s(\rho\|\sigma)\leq E_r^0(\rho\|\sigma)$.
Similarly, we have
$E_r^s(\rho\|\sigma)\geq E_r^0(\rho\|\sigma)$ for any $s< 0$ and $0\leq r<1$.

Thus, in order to prove the
conclusion, it is sufficient to show that
$$\frac{dE^{s}_{r}(\rho\|\sigma)}{ds}\leq 0$$ if $tr(\rho^{r}
\sigma^{1-{r}})>0$ for any $s\neq 0, 0\leq
r<1.$ Note that
\begin{eqnarray}&\ &\frac{dE^{s}_{r}(\rho\|\sigma)}{ds}
=\frac{[-(tr(\rho^{r}\sigma^{1-r}))^{s}\ln
tr(\rho^{r}\sigma^{1-r})]s-
[1-(tr(\rho^{r}\sigma^{1-r}))^{s}]}{(1-r)s^{2}}.
\end{eqnarray}

Let $tr(\rho^{r}\sigma^{1-r})=x$, $f(x)=\frac{x^{s}-1-sx^{s}\ln
x}{(1-r)s^{2}}.$ Then $0<x\leq 1$,
$\frac{dE^{s}_{r}(\rho\|\sigma)}{ds}=\frac{{x}^{s}-1-s{x}^{s}\ln
{x}}{(1-r)s^{2}}$, $f^{'}(x)=\frac{-x^{s-1}\ln x}{1-r}\geq 0,$ so
$f(x)\leq f(1)=0.$ Thus we get
$\frac{dE^{s}_{r}(\rho\|\sigma)}{ds}\leq 0.$

(4) $ \ $ When $r=1$, the conclusion is clear.

When $s=0$, $E_r^s(\rho\|\sigma)=H_r(\rho\|\sigma)$ is a constant function of $s$.
Hence, the conclusion is clear.

When $s\neq 0$, if for some $0\leq
r_0<1$, $tr(\rho^{r_0} \sigma^{1-{r_0}})=0$, then for any $0\leq
r<1$, $tr(\rho^{r} \sigma^{1-{r}})=0$, thus,
$E_r^s(\rho\|\sigma)=\frac{1}{(1-r)s}$ is a convex function of $s$,
hence, the conclusion is true in this case.

When $s\neq 0$, if $tr(\rho^r \sigma^{1-r})>0$ for each $0\leq r<1$. Let
$tr(\rho^{r}\sigma^{1-r})=x$. Then $0< x\leq 1$. Moreover,
$$\frac{d^{2}E^{s}_{r}(\rho\|\sigma)}{ds^{2}}=\frac{-s^{2}{x}^{s}\ln^{2}{x}+2s{x}^{s}\ln {x}+2(1-{x}^{s})}{(1-r)s^{3}}.$$
Let $g(x)=-s^{2}x^{s}\ln^{2}x+2sx^{s}\ln x+2(1-x^{s}).$ Then
$g^{'}(x)=-s^{3}x^{s-1}\ln^{2}x.$ Thus, $g^{'}(x)\leq 0$ if $s>0$, $g^{'}(x)\geq 0$ if $s<0$.
Correspondingly, $g(s)\geq g(1)=0$ if $s>0$, $g(s)\leq g(1)=0$ if $s<0$.
Hence $\frac{d^{2}E^{s}_{r}(\rho\|\sigma)}{ds^{2}}\geq 0$ for $s\neq 0$. The
conclusion is proved finally.

\vskip0.2in

\centerline{\bf References}

\vskip0.2in

\noindent [1]. P. N. Rathie,
 Unified $(r,s)$-entropy and its bivariate measures,
Inf. Sci. {\bf 54}, 23-39, (1991)

\noindent [2]. X. H. Hu and Z. X. Ye,
 Generalized quantum entropy,
J. Math. Phys. {\bf 47}, 023502-1-023502-7, (2006)

\noindent [3]. M. A. Nielsen and I. L. Chuang, {\it Quantum
Computation and Quantum Information.} Cambrige University Press,
Cambrige, (2000)

\noindent [4]. M. Ohya and D. Petz,
 {\it Quantum Entropy and its Use.} Springer-Verlag, Berlin, (1991)

\noindent [5]. S. Furuichi, K. Yanagi and K. Kuriyama,
 Fundamental properties of Tsallis relative entropy,
J. Math. Phys. {\bf 45}, 4868-4877, (2004)

\noindent [6]. S. Furuichi, A note on a parametrically extended
entanglement-measure due to Tsallis relative entropy, INFORMATION.
{\bf 9}, 837-844, (2006)

\noindent [7]. I. J. Taneja, L. Pardo, D. Morales and M. L.
Men$\acute{e}$ndez, On generalized information and divergence
measures and their applications: a brief review,
Q$\ddot{U}$ESTII$\acute{O}$, {\bf 13}, 47-73, (1989)

\noindent [8]. R. G. Douglas, On majorization and range inclusion of
operators in Hilbert space, Proc. Amer. Math. Soc. {\bf 17},
413-416, (1966)

\noindent [9]. A. Wehrl, General properties of entropy, Rev. Mod.
Phys. {\bf 50}, 221-260, (1978)

\noindent [10]. G. Lindblad, Expectations and Entropy Inequalities
for Finite Quantum Systems, Commun. math. Phys. {\bf 39}, 111-119,
(1974)

\noindent [11]. G. Lindblad, Completely Positive Maps and Entropy
Inequalities, Commun. math. Phys. {\bf 40}, 147-151, (1975)

\noindent [12]. M. B. Ruskai, Inequalities for quantum entropy: A
review with conditions for equality, J. Math. Phys. {\bf 43},
4358-4375, (2002)

\noindent [13]. E. H. Lieb, Convex Trace Functions and the
Wigner-Yanase-Dyson Conjecture,  Advan. Math. {\bf 11}, 267-288,
(1973)

\noindent [14]. M. B. Ruskai and F. M. Stillinger, Convexity
inequalities for estimating free energy and relative entropy, J.
 Phys. A {\bf 23}, 2421-2437, (1990)

\end{document}